\documentclass[
 reprint,
 amsmath,amssymb,
 aps, prb
]{revtex4-1}
\usepackage{bm}
\usepackage{SIunits}
\usepackage{graphics}
\usepackage{dcolumn}

\begin{document}


\title{The long-range spin-singlet proximity effect for the Josephson system with single-crystal ferromagnet due to its  band structure features}


\author{M.V. Avdeev}
\email[]{avdeev.maxim.kfu@gmail.com}
\affiliation{Institute of Physics, Kazan Federal University,  Kazan 420008, Russia}

\author{Yu.N. Proshin}
\email[]{yurii.proshin@kpfu.ru}
\affiliation{Institute of Physics, Kazan Federal University,  Kazan 420008, Russia}

\date{\today}

\begin{abstract}
A possible explanation for the long-range proximity effect observed in single-crystalline cobalt nanowires sandwiched between two tungsten superconducting electrodes [Wang, M. \textit{et al}. \textit{ Nat. Phys}. \textbf{6}, 389 (2010)] is proposed. The theoretical model uses properties of a ferromagnet band structure. Specifically, to connect the exchange field with the momentum of quasiparticles the distinction between the effective masses in majority and minority spin-subbands and the Fermi surface anisotropy are considered.  
The derived Eilenberger-like equations  allowed us to obtain a renormalized exchange interaction that is completely compensated for some crystallographic directions under certain conditions. The proposed theoretical model is compared with previous approaches.
\end{abstract}


\maketitle


Recent advances in fabrication and design of layered superconductor (S-) - ferromagnet (-F) structures based on the proximity effect~\cite{deGennes_RMP_1964} have led to significant progress in superconducting spintronics~~\cite{Izyumov_UFN_2002, Buzdin_RMP_2005, Bergeret_RMP_2005,Keizer_nat_2006,Feofanov_2010, Eschrig_PhysTd_2011,Blamire_JoPhCM_2014,Eschrig_RPP_15,Linder_PRB_2010,Singh_PRX_2015,Leksin_PRB_2015}. One of the key questions hotly debated in the past years is an origin of the \textit{long-range} proximity effect. Usually in SF structures, the penetration depth ($ L_{SF} $) of  induced singlet superconducting correlations into the F region is strongly restricted by the exchange field $h$. This tends to align the electron spins in parallel, breaking superconducting Cooper pairs with antiparallel spins~\cite{Izyumov_UFN_2002, Buzdin_RMP_2005}.In conventional ferromagnets such as Co, Fe, {etc.} the penetration depth can be estimated as $L_{SF} \sim\xi_h = \sqrt{D/2h}$ which is about  1--10\,\nano\meter. Here, $D$ is a diffusion constant in the ferromagnet and we assume  $\hbar = k_B = 1$ hereinafter.
This value is much less than the corresponding decay length ($ L_{SN} $) for the nonferromagnetic (N) metals, $L_{SN} \sim \xi_N = \sqrt{D/2\pi T}$, that can reach 0.1--1\,\micro\meter\  at sufficiently low temperatures of $T \ll h$.
Moreover, in contrast to normal metal, Fulde-Ferrell-Larkin-Ovchinnikov-(FFLO-)like superconducting state in ferromagnet has oscillating behavior~\cite{Fulde_PR_1964, Larkin_JETP_1964}.

The \textit{long-range} proximity effect arises if the superconducting correlations in an SF structure become insensitive to the exchange field, and $ L_{SF} $ is comparable to $ L_{SN} $. 
The latter conditions are possible for superconducting triplet correlations with total spin projection $S_z = \pm1$. The triplet type of superconductivity occurs when the exchange field is inhomogeneous~\cite{Bergeret_PRL_2001,Fominov_JETP_2003, Bergeret_RMP_2005,Buzdin_RMP_2005,Houzet_PRB_2007,Linder_PRB_224504_2010,Halterman_PRB_2015,Moor_PRB_2015}. This can be realized in SF multilayers with noncolinear magnetizations in different F layers~\cite{Bergeret_PRB_2001, Fominov_JETP_2003, Fominov_JETPLett_2010, Leksin_PRB_2015, Singh_PRX_2015, Alidoust_AppPhys_2015, Halterman_PRB_2016}, in the presence of  domain walls~\cite{Buzdin_PRB_2011, Fritsch_SUST_2015,Tumanov_LTP_2016} or a spin-active interface~\cite{Cottet_PRB_2005, Linder_PRB_2010}.

Recently, Wang \textit{et al.}~\cite{Wang_NatPhys_2010}  investigated transport properties of single-crystal ferromagnetic cobalt nanowires sandwiched between superconducting tungsten electrodes. 
This was a first observation of a long-range \textit{singlet} proximity effect in clean SFS structures. 
The following features of the cited work were the most striking:   
(a) a zero resistance was detected at the excitation current of about 1\,$\mu$A for a wire length of $L=600$\,\nano\meter\  (the magnitude of the critical current $I_c$ at zero magnetic field for a 40\,\nano\meter-diameter Co nanowire is about $I_c \approx 12$\,\micro\ampere); (b) the Co wires did not contain any magnetic inhomogeneities, and they were single crystal and monodomain.

 Immediately after the appearance of the work~\cite{Wang_NatPhys_2010},  Konschelle \textit{et al.}~\cite{Konschelle_PRB_2010} had suggested an explanation of the observed long-range proximity-induced singlet superconductivity based on one-dimensional (1D) Eilenberger equations~\cite{Eilenberger_ZPhys_1968}. This approach was proposed in well-known work~\cite{Buzdin1982critical}.
 
The authors~\cite{Konschelle_PRB_2010} have obtained that the standard singlet proximity effect becomes long-ranged if the ferromagnet in the SFS structure is considered as a 1D  
ferromagnetic wire in the ballistic transport regime.  
Their estimate for the single-channel critical current was proportional $I_{c0} \sim \cos(L/a_f)$. Note,  this current exhibits undamped strong oscillations on the spin stiffness length of $a_f = \upsilon_F/2h \sim 1$--10\,\nano\meter\  ($\upsilon_F$ is a Fermi velocity). The total critical current $I_c$ is the sum of all $M$ transverse channels ($M \sim 10^5$ for a 40\,\nano\meter-diameter nanowire~\cite{Konschelle_PRB_2010}). This total current is very sensitive to small fluctuations of $L$, and  $I_c$ should disappear after averaging $I_c \sim M\langle I_{c0}\rangle _{\delta L} \rightarrow 0$. In reality, the contributions from different channels are not strictly coherent due to $\langle \delta L\rangle = 0$, $\langle (\delta L)^2\rangle \sim a_f^2 $. 

  Another model has been proposed afterwards in works~\cite{Melnikov_PRL_2012, Samokhvalov_UFN_2016}  where  the long-range triplet superconducting correlations were associated with the spin-orbit interaction in F nanowires. 
 In this case, the effective exchange field depends on the quasiparticle momentum and it strongly affects the phase gain along the trajectories. 
 The long-range contributions to 
 the supercurrent are due to the modulation of the momentum-dependent exchange field along the quasiparticle trajectories. 
 It is important that the lengths of paths between successive reflections should coincide, then the corresponding phases compensate each other.
 For an explanation of the experimental data~\cite{Wang_NatPhys_2010}, the authors~\cite{Melnikov_PRL_2012, Samokhvalov_UFN_2016} used a two-dimensional model of a ferromagnet nanowire 
 with multiple ideal reflections from the boundaries. 
 Furthermore, in works~\cite{Bergeret_PRL_2013,Bergeret_PRB_2014} Bergeret and Tokatli showed analogy between the spin-diffusion process in normal metals and the generation of the triplet correlations in a diffusive superconducting structure in the presence of a spin-orbit coupling. From this analogy it turns out that the spin-orbit coupling is an additional source for the long-range triplet components besides the magnetic inhomogeneities.
 
 At last, in the work~\cite{Melnikov_PRL_2016} Mel'nikov and Buzdin have demonstrated that giant mesoscopic fluctuations arising in dirty ferromagnetic wires can also result in a long-order Josephson current, but the value of the effect drastically changes ``from-sample-to-sample''. 
  
Our approach is based on the known physical fact that the effective masses of the conduction electrons for spin bands $(1/m^\alpha)_{ij} = \partial^2 \varepsilon_\alpha(\mathbf{k})/\partial k_i \partial k_j$  are generally different in real ferromagnets~\cite{Batallan_PRB_1975, Monastra_PRL_2002, Schafer_PRB_2005}.  Here $\alpha = \uparrow(\downarrow)$ labels spins in the majority (minority) spin-subband, respectively. Indeed, this feature can lead to a compensation of the total momentum of the Cooper pair in a ferromagnet. It is easy to understand within the simple picture of the  FFLO pairing mechanism~\cite{Fulde_PR_1964,Larkin_JETP_1964} with total momentum $\mathbf{q}$ of the pair ($q$ is much less than the Fermi momentum $k_F$). In ferromagnet the momentum $\mathbf{q}$ is obtained from the condition 
$
(\mathbf{k}_F + \mathbf{q}/2)^2/2m^\uparrow - h = (-\mathbf{k}_F + \mathbf{q}/2)^2/2m^\downarrow + h.
$
It follows immediately that $\mathbf{q}\,\mathbf{k}_F/2M\approx h - \eta\,{k}_F^2/2M$, where $M = 2m^\uparrow m^\downarrow/(m^\downarrow + m^\uparrow)$ and mismatch parameter $\eta = (m^\downarrow - m^\uparrow)/(m^\downarrow + m^\uparrow)$. Thus the total momentum of the FFLO-like pair completely vanishes at $\eta \approx h/E_F \ll 1$, where $E_F$ is the Fermi energy. It leads to a long-range spatial extent of the induced superconductivity in a ferromagnetic nanowire.

In contrast to previous theoretical works~\cite{Konschelle_PRB_2010,Melnikov_PRL_2012}, we focus on a case of three-dimensional (3D) nanowires. We would like to stress that the Co nanowires with diameters $d$ of 40 and 80\,nm were investigated in experiment~\cite{Wang_NatPhys_2010}, and these values  are considerably larger than the bare spin stiffness length, $d \gg a_f$. As a consequence, the model of a 3D nanowire is the most relevant one to the experimental setup~\cite{Wang_NatPhys_2010}. 
However, our approach can be applied for arbitrary dimension.
  
In this Rapid Communication we propose a theory  
of the singlet long-range proximity effect in  single-crystal ferromagnetic nanowires based on the following key points: 
(a)~The conduction electrons have different effective masses in the majority and minority spin subbands; 
(b)~the Josephson transport in single-crystal nanowires takes place in the ballistic regime (the clean case);
(c)~the Fermi surface in the ferromagnet is anisotropic.

The anisotropic dispersion relation  
supposed for a hexagonal close-packed single-crystal cobalt nanowire is
 \begin{equation*}
 \varepsilon_\alpha(\mathbf{k}) = \frac{k_x^2}{2m_{\perp}^{\alpha}} + \frac{k_y^2}{2m_{\perp}^{\alpha}} + \frac{k_z^2}{2m_{\parallel}^{\alpha}} - h({\sigma}_3)_{\alpha\alpha},
 \end{equation*}
 where $\hat{\sigma}_3$ is the third Pauli matrix. The Matsubara-Green's  function $ \hat{G}  $
  satisfies the equations,
 \begin{eqnarray}\label{eq:green1}
 \hat{G}^{-1}(\mathbf{k} + \mathbf{q}/2,\omega)\hat{G}(\mathbf{k},\mathbf{q}, \omega) = \delta(\mathbf{q}), \\\label{eq:green2}
 \hat{G}(\mathbf{k}, \mathbf{q},\omega)\hat{G}^{-1}(\mathbf{k}-\mathbf{q}/2, \omega) = \delta(\mathbf{q}),
 \end{eqnarray} 
where $\omega = \pi T(2n+1)$ is the Matsubara frequency, and $\hat{G}^{-1}$ in a ferromagnet nanowire  
has the form
\begin{equation}
\begin{aligned}\label{eq:invGreen}
\hat{G}^{-1}(\mathbf{k}) &= \begin{pmatrix}
i\omega  - \varepsilon_\uparrow(\mathbf{k}) +\mu & 0\\
0 & -i\omega - \varepsilon_\downarrow(-\mathbf{k})+\mu
\end{pmatrix} \\
&=  \left[i\omega + h_{eff}(\mathbf{k})\right]\hat{\sigma}_3 - \left[E(\mathbf{k}) - \mu\right]\hat{\sigma}_0,
\end{aligned}
\end{equation}
where $\mu$ is the chemical potential, $\hat{\sigma}_0$ is the unit matrix and the superconducting order parameter $\Delta$ is assumed to be zero in the ferromagnet. 
It is important to note that the mismatch between $m_\downarrow$ and  $m_\uparrow$ leads to an appearance of  \emph{effective} exchange interaction $h_{eff}(\mathbf{k})=(\varepsilon_\downarrow(-\mathbf{k}) - \varepsilon_\uparrow(\mathbf{k}))/2$ and \emph{effective} paramagnetic dispersion  $E(\mathbf{k}) = (\varepsilon_\uparrow(\mathbf{k}) + \varepsilon_\downarrow(-\mathbf{k}))/2$
in~\eqref{eq:invGreen}, and they become dependent on the momentum as follows:
\begin{equation}
\begin{aligned}\label{eq:exch_eff}
h_{eff}(\mathbf{k}) &= h - \eta_\perp\left(\frac{k_x^2}{2M_\perp} + \frac{k_y^2}{2M_\perp} \right) -\eta_\parallel\frac{k_z^2}{2M_\parallel},\\
E(\mathbf{k}) &= \frac{k_x^2}{2M_\perp} + \frac{k_y^2}{2M_\perp} + \frac{k_z^2}{2M_\parallel},
\end{aligned}
\end{equation}
where the mismatch parameters $\eta_{\parallel}$, $\eta_\perp$ and reduced masses $M_{\parallel}$, $M_\perp$ are defined as
\begin{equation}
\begin{aligned} 
\eta_{\parallel} &= \frac{m_{\parallel}^\downarrow - m_{\parallel}^\uparrow}{m_{\parallel}^\downarrow + m_{\parallel}^\uparrow},\quad
\eta_{\perp} = \frac{m_{\perp}^\downarrow - m_{\perp}^\uparrow}{m_{\perp}^\downarrow + m_{\perp}^\uparrow},\\
M_{\parallel} &=\frac{2m_{\parallel}^\uparrow m_{\parallel}^\downarrow}{m_{\parallel}^\uparrow + m_{\parallel}^\downarrow},\quad M_{\perp} =\frac{2m_{\perp}^\uparrow m_{\perp}^\downarrow}{m_{\perp}^\uparrow + m_{\perp}^\downarrow}. 
\end{aligned}
\end{equation}
It is easy to see that in the isotropic case when $E(\mathbf{k})=k^2/2m$, then  $\eta_{\parallel}=\eta_\perp=\eta$. In the limit case of $\eta =0$ the \emph{effective} exchange field coincides with the bare one $h_{eff}(\mathbf{k}) \equiv h$.

Furthermore, subtracting  Eq.~\eqref{eq:green2} from  Eq.~\eqref{eq:green1} and passing into the coordinate representation ($\mathbf{q}\rightarrow -i\nabla_{\mathbf{R}}$) in the usual manner, we obtain a quasiclassical Eilenberger-like equation~\cite{Eilenberger_ZPhys_1968}  in a ferromagnetic nanowire, 
\begin{equation}
\begin{aligned}
&i\bm{\upsilon}_0(\mathbf{k_0})\nabla_{\mathbf{R}} \hat{\mathcal{G}}+[(i\omega + h_{eff}(\mathbf{k}_0))\hat{\sigma}_3, \, \hat{\mathcal{G}}] = 0,\\
&\hat{\mathcal{G}}(\mathbf{R},\mathbf{k}_0,\omega) = \int \frac{d\xi}{2\pi}\,\hat{G}(\mathbf{R},\xi,\mathbf{k}_0,\omega)= \frac12\begin{pmatrix}
-i g & f \\
-f^\dagger & i g
\end{pmatrix},
\end{aligned}
\end{equation} 
where the momentum $\mathbf{k}_0$ is defined as $E_F = E(\mathbf{k}_0)$, the corresponding 
velocity is $\bm{\upsilon}_0(\mathbf{k}_0) = \nabla E(\mathbf{k}_0) = (\bm{\upsilon}_{F\uparrow} + \bm{\upsilon}_{F\downarrow})/2$ and $\xi = E(\mathbf{k}) - \mu$. 
\begin{figure}[t]
	\includegraphics{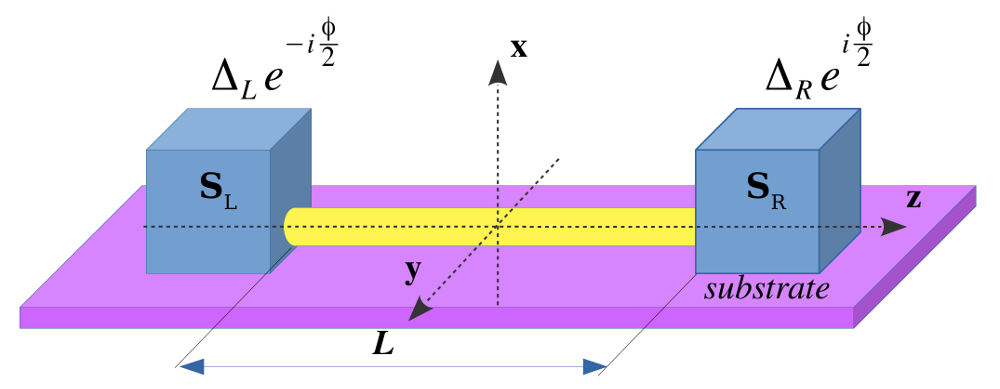}
	\caption{\label{fig:geometry} Schematic of the Josephson junction with a ferromagnetic single-crystal nanowire of length $L$ sandwiched between superconducting electrodes.}
\end{figure}
\begin{figure*}[ht]
	\centering
	\includegraphics{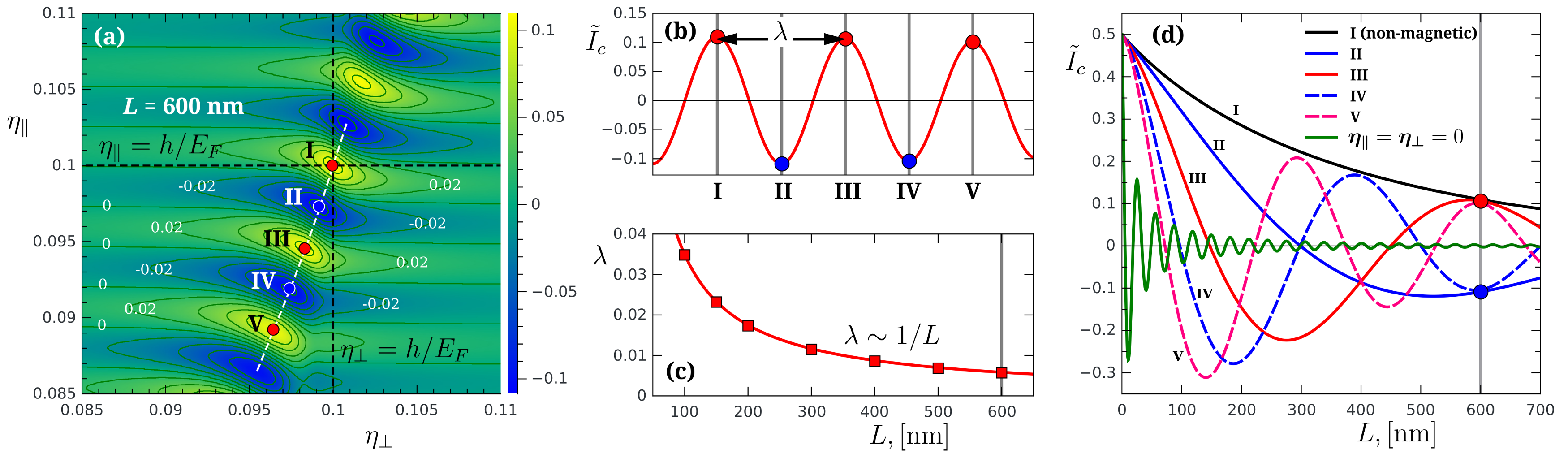}
	\caption{\label{fig:critCurrent}  (a)~The map of the reduced critical current $\tilde{I}_c$ 
as a function of  the	
		mismatch parameters $\eta_\parallel$ and $\eta_\perp$ for a fixed nanowire length of $L=600$\,nm. The non-magnetic case corresponds to  point (I) at $\eta_\parallel = \eta_\perp = h/E_F = 0.1$. (b)~The $\tilde{I}_c$ oscillations 
	along the 
	path passing through peaks I-V at $L=600$\,nm.    
		(c)~The dependence of the peak period $\lambda$ versus nanowire length. 
		(d)~The reduced critical current $\tilde{I}_c$ as a function of nanowire length. 
		Lines I-V are consistent with  points I-V in panel (a), i.e., corresponding parameters are optimal 
		for length $L=600$ nm.}
\end{figure*} 
The current density in the quasiclassical approach can be expressed as
\begin{equation}
\mathbf{j} = -ie T\sum_\omega \oint_{E_F}\bm{\upsilon}_0(\mathbf{k}_0) g_\omega(\mathbf{k_0}, \mathbf{R})\frac{ds}{|\nabla E(\mathbf{k}_0)|(2\pi)^2},
\end{equation}
where the integration is performed 
over the Fermi surface.
Let us now consider the Josephson transport through a single-crystal ferromagnet nanowire according to the experimental setup~\cite{Wang_NatPhys_2010}. 
Thus, a wire of 
length $L$ and cross-sectional $S$ is placed between the left and the right superconducting electrodes (S$_{L(R)}$) located at $z = \pm L/2$ as shown in Fig.~\ref{fig:geometry}. 
We introduce the angle $\theta$ between the momentum $\mathbf{k_0}$ and the $z$ axis so that
\begin{equation}
\begin{aligned}
E_F &= k_0^2\left(\frac{\cos^2\theta}{2M_\parallel} + \frac{\sin^2\theta}{2M_\perp}\right),\\
h_{eff}(\theta) &= h - k_0^2\left(\eta_{\parallel}\frac{\cos^2\theta}{2M_\parallel} + \eta_{\perp}\frac{\sin^2\theta}{2M_\perp}\right),\\
\upsilon_{0z}(\theta) &= \frac{k_{0z}}{M_\parallel} =  \sqrt{\frac{2E_F}{M_\parallel}}\frac{\cos\theta}{\sqrt{\cos^2\theta + \frac{M_\parallel}{M_\perp}\sin^2\theta}}.
\end{aligned}
\end{equation}
Thus, the Josephson supercurrent flowing across the nanowire is given by
\begin{equation}\label{eq:Josephson_current}
 \begin{aligned}
I &= -ie S T\sum_{\omega}\oint_{E_F} \tau_z(\theta)\,g(z,\theta,\omega)\, \frac{ds}{(2\pi)^2},\\
\tau_z(\theta) &= \upsilon_{0z}/|\bm{\upsilon}_0| = \frac{\cos\theta}{\sqrt{\cos^2\theta + (M_\parallel/M_\perp)^2\sin^2\theta}},\\
ds &= k_0^2\,d\Omega = \frac{2M_\parallel E_F}{\cos^2\theta + ({M_\parallel}/{M_\perp})\sin^2\theta}d\Omega,
\end{aligned}
\end{equation} 
where $d\Omega$ is a solid angle element.
The anomalous Green's functions $f$, $f^\dagger$ in the ferromagnet satisfy  the following equations  
\begin{equation}
\begin{aligned}\label{eq:Eilenberger}
\upsilon_{0z}(\theta)&\frac{\partial}{\partial z} f + 2 f(\omega - i h_{eff}(\theta)) = 0,\\
-\upsilon_{0z}(\theta)&\frac{\partial}{\partial z} f^\dagger + 2 f^\dagger(\omega - i h_{eff}(\theta)) = 0, \\
\end{aligned}
\end{equation}
with the rigid boundary conditions ($\cos\theta > 0$) 
\begin{equation}
\begin{aligned}
f(-L/2) = \frac{\Delta_L}{|\omega|} e^{-i\phi/2}, \,\,\, f^\dagger(L/2) = \frac{\Delta_R}{|\omega|} e^{-i\phi/2}, 
\end{aligned}
\end{equation}
which are valid when the superconducting electrodes are much thicker than the nanowire's cross section. 
Using the normalization condition $g^2 + f^\dagger f = 1$, we obtain 
$$
g\approx \mathrm{sign}(\omega)\left(1 - \frac{1}{2}f^\dagger f\right)
$$
and 
the Josephson supercurrent~\eqref{eq:Josephson_current} is transformed to the form
\begin{equation}
I = I_c\sin\phi, \quad {I_c} = 2S\frac{e M_\parallel E_F \Delta_L \Delta_R}{\pi^3 T} \tilde{I}_{c}(L),
\end{equation}
where the reduced critical current $\tilde{I}_c$ defines the spatial extent of the induced superconductivity in nanowire as follows:
\begin{equation}
\begin{aligned}\label{eq:critical_current}
\tilde{I}_{c}(L) &= \int_{0}^{1}  \frac{\cos\theta\,d(\cos\theta)}{\sqrt{\cos^2\theta + (M_\parallel/M_\perp)^2\sin^2\theta}}\times\\
&\times\frac{1}{\cos^2\theta + ({M_\parallel}/{M_\perp})\sin^2\theta}\times\\ 
&\times \exp\left(-\frac{2\pi T L}{\upsilon_{0z}(\theta)} \right) \cos\left(\frac{2 h_{eff}(\theta)L}{\upsilon_{0z}(\theta)}\right).
\end{aligned}
\end{equation}
Note that 
the critical current for a 1D case  
can be written in our theory framework as  
$I_c\sim \exp\left(-{2\pi T L}/{\upsilon_{0z}(0)} \right) \cos\left({2 h_{eff}(0)L}/{\upsilon_{0z}(0)}\right)$ which agrees with the results of previous studies~\cite{Konschelle_PRB_2010} in the limiting case when the band masses are equal (i.e., when $\eta_\parallel = \eta_\perp = 0$ and hence $h_{eff} = h$). If $\eta_\parallel = \eta_\perp =  h/E_F$, then $h_{eff}(0) = 0$, and we obtain a new important 
limiting case of f \emph{normal} non-ferromagnetic nanowire.  

For numeric estimations we assume that both mismatch parameters are small $\eta_\parallel, \eta_\perp \ll 1$ and ratio $M_\parallel/M_\perp \approx 1$.  We also set the bare spin stiffness length of $a_{fz} = \upsilon_{0z}(0)/2h = 5$\,\nano\meter, coherence length of $\xi_{fz} = \upsilon_{0z}(0)/2\pi T = 600$\,\nano\meter\ and ratio $h/E_F = 0.1$ for the Co nanowire. 
The map of the reduced critical current $\tilde{I}_c$ as a function of both mismatch parameters $\eta_\parallel$ and $\eta_\perp$ for the fixed nanowire length of $L=600$\,nm is shown in Fig.~\ref{fig:critCurrent}(a).  Point I ($\eta_\parallel = \eta_\perp = h/E_F$), as mentioned above, corresponds to the non magnetic case where the effective exchange field is completely compensated [$h_{eff} = 0$, see Eq.~\eqref{eq:Eilenberger}] for all trajectories. 

It is clearly seen 
that $\tilde{I}_c$ has multiple peaks with a periodic sign-change behavior. The points where $\tilde{I}_c > 0$ (I, III, V, {etc}.) and $\tilde{I}_c <0$ (II, IV, {etc}.) correspond to the so-called $0$ and $\pi$ states of the Josephson junction, respectively. The appearance of multiple peaks is a consequence of the fact that the wave functions of the Cooper pairs in the ferromagnet have an \emph{effective} momentum $q_{z}(\theta) \approx h_{eff}(\theta)/\upsilon_{0z}(\theta)$ and oscillate  along the trajectory. As a result, the contribution from all quasi-classical trajectories  between the superconducting electrodes leads to 
an unusual interference pattern. Figure~\ref{fig:critCurrent}(b) shows the slice $\tilde{I}_c$  
along the I-V 
line.  
The distance $\lambda$ between neighboring peaks of the same sign is depicted as a function of the nanowire length $L$ in Fig.~\ref{fig:critCurrent}(c). The function $\lambda(L)$ shows a sufficiently slow monotonic behavior as $\lambda \sim 1/L$ [the fit of the red solid line in Fig.~\ref{fig:critCurrent}(c)]. 
If the mismatch parameters $\eta_\parallel$, $\eta_\perp$ take the values close to the line along peaks [the white dashed line in Fig. 2(a)], then we observe slow detectable oscillations of the critical current $I_c$  with a change in the nanowire length.
For a clear visualization the five spatial curves $I_c(L)$ are presented in Fig.~2(d) at set points ($\eta_\parallel$, $\eta_\perp$) that correspond to I-V
peaks at $L=600$\,nm [see Fig.~\ref{fig:critCurrent}(a)]. Note that the function  $\tilde{I}_c(L)$ monotonically decays for the non-magnetic regime (curve I with $\eta_\parallel=\eta_\perp = h/E_F$) on a scale about of the coherence length $\xi_{fz}$, that is in agreement with the  physical picture of the proximity effect for the SNS Josephson junction. 
 
The oscillating behavior $\tilde{I}_c(L)$ arises even at a small deviation of mismatch parameters $\eta_\parallel$, $\eta_\perp$ from  point I. For example, in the range of 0--600\,nm,   curves II and III exhibit 0-$\pi$ and 0-$\pi$-0-crossovers, respectively, and the period of oscillations decreases with each subsequent curve (IV, V, {etc}.). 
For comparison, the solid green curve in Fig.~\ref{fig:critCurrent}(d) reproduces the limiting case of $\eta_\perp = \eta_\parallel = 0$ when the majority and minority band masses are equal $m_\uparrow = m_\downarrow$. This equality is common for standard models of the proximity effect in SF structures~\cite{Izyumov_UFN_2002,Buzdin_RMP_2005,Bergeret_RMP_2005}. As is clearly seen in Fig.~2(d) the singlet long-range Josephson current does not arise in this limiting case. We note that within our theory the inequality $m^\downarrow_{\parallel(\perp)} > m^\uparrow_{\parallel(\perp)}$ (and hence $\eta_\perp, \eta_\parallel > 0$) gives rise to a singlet long-range proximity effect.
We also see that  $\tilde{I}_c$ has noticeable stability and the critical current varies weakly with a relatively large change in the nanowire length of $\delta L \sim 100$\,nm in contrast to the case when $m_\uparrow = m_\downarrow$ [the solid green line in Fig.~\ref{fig:critCurrent}(d)].

\textit{To summarize}, we propose a singlet mechanism of the long-range proximity effect in superconductor-ferromagnet structures. Our approach is based on a simple physical picture where the spin-subband electron masses are different. The energy dispersion anisotropy leads to the appearance of a set of points ($\eta_\parallel, \eta_\perp$) for which a long-range Josephson effect is possible.
Note that in the isotropic case, only a sole mismatch parameter is possible.
In our case, the region of parameters where the long-range effect is noticeable, is sufficiently broad. The proposed mechanism gives a possible explanation of the experiment by Wang \textit{et al.}~\cite{Wang_NatPhys_2010}. 
As a final note, the considered approach is not applicable for polycrystalline samples.

\begin{acknowledgments}
The Rapid Communication was supported by the subsidy of the Ministry of Education and Science of the Russian Federation (Grant No. 3.2166.2017) allocated to Kazan Federal University for performing the project part of the state assignment in the area of scientific activities. The authors are grateful Dr. S.K. Saikin  for valuable advice and comments. MVA is also thankful to the RFBR (Grant No. 16-02-01016) for partial support.
\end{acknowledgments}

\bibliography{bibliography}

\end{document}